\documentclass{article}
\usepackage{amssymb}
\usepackage{amsfonts}
\usepackage{amsmath}

\input{tcilatex}

\begin{document}

\title{Charged coherent states related to $su_{q}(2)$ covariance\thanks{%
The project supported by National Natural Science Foundation of China under
Grant No. 10075042}}
\author{Yun Li\thanks{%
Email: yunlee@mail.ustc.edu.cn} and Sicong Jing\thanks{%
Email: sjing@ustc.edu.cn} \\
\textit{Department of Modern Physics, }\\
\textit{\ University of Science and Technology of China,}\\
\textit{\ Hefei, Anhui 230026, P.R.China }}
\maketitle

\begin{abstract}
A new kind of $q$-deformed charged coherent states is constructed in Fock
space of two-mode $q$-boson system with $su_{q}(2)$ covariance and a
resolution of unity for these states is derived. We also present a simple
way to obtain these coherent states using state projection method.
\end{abstract}

With the study of solutions of the Yang-Baxter equation, quantum groups and
algebras have been extensively developed by Jimbo\cite{1} and Drinfeld\cite%
{2}. To realize the Jordan-Schwinger mapping of the quantum algebra $%
su_{q}\left( 2\right) $, Biedenharn and Macfarlane\cite{3}\cite{4}
introduced a kind of $q$-deformed harmonic oscillators. They also discussed
some properties of these oscillators, for example, the Fock space structure
and the $q$-deformed coherent states for such oscillator. And Sun and Fu\cite%
{5} gave the the $q$-deformed boson realization of the quantum algebra $%
su_{q}\left( n\right) $ constructed in terms of this kind of deformed
oscillators. But the oscillators in their discussions are all mutually
commuting. To our knowledge, in 1989, Pusz and Woronowicz first introduced
multimode $q$-deformed oscillators\cite{6} which are not mutually commuting
but satisfy the following relations

\begin{eqnarray}
a_{i}^{\dag }a_{j}^{\dag } &=&\sqrt{q}a_{j}^{\dag }a_{i}^{\dag }\text{ \ \ \ 
}\left( i<j\right) \text{ \ }\left( i,j=1,\cdots ,n\right)  \notag \\
a_{i}a_{j} &=&\frac{1}{\sqrt{q}}a_{j}a_{i}\text{ \ }\left( i<j\right)  \notag
\\
a_{i}a_{j}^{\dag } &=&\sqrt{q}a_{j}^{\dag }a_{i}  \label{1} \\
a_{j}a_{i}^{\dag } &=&\sqrt{q}a_{i}^{\dag }a_{j}  \notag \\
a_{i}a_{i}^{\dag } &=&1+qa_{i}^{\dag }a_{i}+\left( q-1\right) \underset{k>i}{%
\sum }a_{k}^{\dag }a_{k}\text{ .}  \notag
\end{eqnarray}%
Kulish and Damaskinsky pointed that these coupled multimode $q$-deformed
oscillators can be expressed in terms of independent $q$-deformed harmonic
oscillators\cite{7}. Wess and Zumino\cite{8} developed a differential
calculus on the quantum hyperplane covariant with respect to the action of
the quantum group $GL_{q}(n)$. Recently Chung studied multiboson realization
of the two-mode $q$-boson algebra relations with $su_{q}(2)$ covariance\cite%
{9}. The algebra relations are given by

\begin{align}
a^{\dagger }b^{\dagger }& =\sqrt{q}b^{^{\dagger }}a^{\dagger }  \notag \\
ab& =\frac{1}{\sqrt{q}}ba  \notag \\
ab^{\dagger }& =\sqrt{q}b^{\dagger }a  \label{2} \\
ba^{^{\dagger }}& =\sqrt{q}a^{\dagger }b  \notag \\
aa^{\dagger }& =1+qa^{\dagger }a+(q-1)b^{\dagger }b  \notag \\
bb^{\dagger }& =1+qb^{\dagger }b\text{ .}  \notag
\end{align}%
In fact, these two-mode $q$-boson algebra relations are exactly the eq$(1)$
for $n=2$ case. Comparing $(2)$ with Wess-Zumino's differential calculus
relations on the covariant quantum plane

\begin{align}
xy& =\sqrt{q}yx  \notag \\
\partial _{x}\partial _{y}& =\frac{1}{\sqrt{q}}\partial _{y}\partial _{x} 
\notag \\
\partial _{x}y& =\sqrt{q}y\partial _{x}  \label{3} \\
\partial _{y}x& =\sqrt{q}x\partial _{y}  \notag \\
\partial _{x}x& =1+qx\partial _{x}+(q-1)y\partial _{y}  \notag \\
\partial _{y}y& =1+qy\partial _{y}\text{ ,}  \notag
\end{align}%
we can easily find that they are the same thing in sense of Bargmann
representation\cite{10} according to the corresponding relations $(a^{\dag
}\leftrightarrow x$, $b^{\dag }\leftrightarrow y$ and $a\longleftrightarrow
\partial _{x}$, $b\leftrightarrow \partial _{y})$. We also gave the the
structure of Fock space for the coupled two-mode $q$-boson system and
discussed its simply application to Jordan-Schwinger realization of quantum
algebra $su_{q}(2)$ and $su_{q}(1,1)$\cite{11}.

In this letter, we construct a new kind of $q$-deformed charged coherent
states in the Fock space of the coupled two-mode $q$-boson oscillators (we
emphasize that these coherent states are not so-called $su_{q}(2)$ coherent
states but they are related to $q$-boson algebra with $su_{q}(2)$
covariance) and show a resolution of unity for these states. We also discuss
a simple way to obtain these states using projection method. Bhaumik \textit{%
et al}\cite{12}\textit{\ }discussed the undeformed and also independent case%
\textit{.} We use the method developed by them.

To begin with , we recall the Fock space representation of $(2)$. We take $%
a^{\dag },a$ and $b^{^{\dag }},b$ as $q$-deformed creation and destruction
boson operators, and define the number operators $N_{a,}$ $N_{b}$ following
by \ \ \ 

\begin{eqnarray}
b^{\dagger }b &=&\left[ N_{b}\right]  \notag \\
q^{-N_{b}}a^{\dagger }a &=&\left[ N_{a}\right]  \label{4}
\end{eqnarray}%
where%
\begin{equation*}
\left[ N\right] =\dfrac{q^{N}-1}{q-1}\text{ .}
\end{equation*}%
Using the algebraic relation$(2)$, it is easy to verify

\begin{eqnarray}
\left[ N_{a},a\right] &=&-a,\ \ \ \left[ N_{b},b\right] =-b,  \notag \\
\left[ N_{a},b\right] &=&0,\ \ \ \ \left[ N_{b},a\right] =0,  \label{5}
\end{eqnarray}%
and $\left[ N_{a},N_{b}\right] =0$. Let $\left\vert 0,0\right\rangle $ be
the ground state satisfying

\begin{eqnarray}
a\left\vert 0,0\right\rangle &=&0,\ b\left\vert 0,0\right\rangle =0  \notag
\\
\ N_{i}\left\vert 0,0\right\rangle &=&0,\ (i=a,b)  \label{6}
\end{eqnarray}%
and $\left\{ \left\vert n,m\right\rangle \mid n,m=0,1,2...\right\} $ be the
set of the orthogonal number eigenstates

\begin{eqnarray}
N_{a}\left\vert n,m\right\rangle &=&n\left\vert n,m\right\rangle  \notag \\
N_{b}\left\vert n,m\right\rangle &=&m\left\vert n,m\right\rangle  \label{7}
\\
\left\langle n,m\mid n^{^{\prime }},m^{^{\prime }}\right\rangle &=&\delta
_{nn^{^{\prime }}}\delta _{mm^{^{\prime }}}\text{ .}  \notag
\end{eqnarray}%
From the algebra$(2)$, the Fock space representation for the $q$-bosons $a$
and $b$ is given by

\begin{eqnarray}
a\left\vert n,m\right\rangle &=&\sqrt{q^{m}\left[ n\right] }\left\vert
n-1,m\right\rangle ,\ \ \ \ \ \ \ b\left\vert n,m\right\rangle =\sqrt{\left[
m\right] }\left\vert n,m-1\right\rangle  \notag \\
a^{\dagger }\left\vert n,m\right\rangle &=&\sqrt{q^{m}\left[ n+1\right] }%
\left\vert n+1,m\right\rangle ,\ \ b^{\dagger }\left\vert n,m\right\rangle =%
\sqrt{\left[ m+1\right] }\left\vert n,m+1\right\rangle \text{ .}  \label{8}
\end{eqnarray}%
The general number eigenstate\ $\left\vert n,m\right\rangle $ is obtained by
applying $b^{\dagger }$ m times after applying $a^{\dagger }$ n times on the
ground state$\left\vert 0,0\right\rangle $

\begin{equation}
\left\vert n,m\right\rangle =\dfrac{\left( b^{\dagger }\right) ^{m}\left(
a^{\dagger }\right) ^{n}}{\sqrt{\left[ n\right] !\left[ m\right] !}}%
\left\vert 0,0\right\rangle  \label{9}
\end{equation}%
where

\begin{equation*}
\left[ n\right] !=\left[ n\right] \left[ n-1\right] \cdots \left[ 2\right] %
\left[ 1\right] ,\ \ \ \ \left[ 0\right] !=1\text{ .}
\end{equation*}

Because of the noncommutative character of $a,b$ showing in$(2)$, they have
not common eigenstate-coherent states. But using the above number operators $%
N_{a}$ and $N_{b}$, we may define the charge operators given by

\begin{equation}
Q=N_{a}-N_{b}  \label{10}
\end{equation}%
which means that each of the $a$ quanta possesses a charge $^{\prime
}+1^{\prime }$, and each of the $b$ quanta a charge $^{\prime }-1^{\prime }$%
. From$(2)$, we can easily find

\begin{equation}
\left[ Q,ab\right] =\left[ Q,ba\right] =\left[ ab,ba\right] =0\text{ .}
\label{11}
\end{equation}%
Hence, \ the operator $Q$, $ab$, and $ba$ may have common eigenstates, which
is called $q$-deformed charged coherent state\cite{12}. We denote this
coherent state as $\left\vert z,e\right\rangle $ ($z$ is a complex number
and $e$ an integer) with the requirements that

\begin{equation}
Q\left\vert z,e\right\rangle =e\left\vert z,e\right\rangle \ \ \ \ \ \ \
ab\left\vert z,e\right\rangle =z\left\vert z,e\right\rangle \text{ .}
\label{12}
\end{equation}

To obtain an explicit expression for this coherent state, we consider the
following expansion

\begin{equation}
\ \ \left\vert z,e\right\rangle =\overset{\infty }{\underset{n,m=0}{\sum }}%
C_{nm}(z)\left\vert n,m\right\rangle \text{ .}\   \label{13}
\end{equation}%
Since $\left\vert z,e\right\rangle $ is an eigenstate of $Q,$ for $%
e\geqslant 0$, we obtain%
\begin{equation}
\left\vert z,e\right\rangle =\overset{\infty }{\underset{m=0}{\sum }}%
C_{m+e,m}(z)\left\vert m+e,m\right\rangle \text{ .}  \label{14}
\end{equation}%
Substituting this expression into the second equation in $\left( 12\right) $%
, using $\left( 8\right) $, we obtain%
\begin{equation}
C_{m+e.m}\left( z\right) =C_{e,0}\dfrac{\sqrt{\left[ e\right] !}z^{n}}{\sqrt{%
q^{m\left( m-1\right) /2}\left[ m\right] !\left[ m+e\right] !}}\text{ ,}
\label{15}
\end{equation}%
so the charged coherent state is given by%
\begin{equation}
\left\vert z,e\right\rangle =N_{e}\overset{\infty }{\underset{m=0}{\sum }}%
\dfrac{z^{n}}{\sqrt{q^{m\left( m-1\right) /2}\left[ m\right] !\left[ m+e%
\right] !}}\left\vert m+e,m\right\rangle  \label{16}
\end{equation}%
with the normalization constant

\begin{equation}
N_{e}^{-2}=\overset{\infty }{\underset{m=0}{\sum }}\ \dfrac{\left\vert
z\right\vert ^{2n}}{q^{m\left( m-1\right) /2}\left[ m\right] !\left[ m+e%
\right] !}\text{ .}  \label{17}
\end{equation}

To resolve the unity for these coherent states, we introduce the $q$%
-analogue of integral defined by Jackson\cite{13}\cite{14}\cite{15}. We
hereby give a few words about this integral. The $q$-derivative is defined as

\begin{equation}
\dfrac{df\left( x\right) }{d_{q}x}=\dfrac{f\left( qx\right) -f\left(
x\right) }{\left( q-1\right) x}  \label{18}
\end{equation}%
and the corresponding integral is given by

\begin{equation}
\int f\left( x\right) d_{q}x=\left( 1-q\right) \overset{\infty }{\underset{%
l=0}{\sum }}q^{l}xf\left( q^{l}x\right) +\mathit{const}\text{ .}  \label{19}
\end{equation}%
If $a$ is finite, we have

\begin{equation}
\int_{0}^{a}f\left( x\right) d_{q}x=\left( 1-q\right) \overset{\infty }{%
\underset{l=0}{\sum }}q^{l}xf\left( q^{l}a\right)  \label{20}
\end{equation}%
and if $a$ is infinitive, we have\ \ \ 
\begin{equation}
\int_{0}^{\infty }f\left( x\right) d_{q}x=\left( 1-q\right) \overset{\infty }%
{\underset{l=-\infty }{\sum }}q^{l}xf\left( q^{l}\right) \text{ .}
\label{21}
\end{equation}%
The $q$-integration by parts formula is%
\begin{equation}
\ \dfrac{d}{d_{q}}\left( f\left( x\right) g\left( x\right) \right) =\left( 
\dfrac{d}{d_{q}}f\left( x\right) \right) g\left( qx\right) +f\left( x\right)
\left( \dfrac{d}{d_{q}}g\left( x\right) \right)  \label{22}
\end{equation}%
\ and%
\begin{equation}
\int_{0}^{a}f\left( x\right) \left( \dfrac{d}{d_{q}}g\left( x\right) \right)
d_{q}x=f\left( x\right) g\left( x\right) \mid
_{x=0}^{x=a}-\int_{0}^{a}\left( \dfrac{d}{d_{q}}f\left( x\right) \right)
g\left( qx\right) d_{q}x\text{ .}  \label{23}
\end{equation}%
It is well known that Exton\cite{16} defines a family of $q$-exponential
functions by\ \ \ \ \ \ \ \ \ \ \ \ \ \ 
\begin{equation}
E\left( q,\lambda ;x\right) =\overset{\infty }{\underset{n=0}{\sum }}\dfrac{%
x^{n}q^{\lambda n\left( n-1\right) }}{\left[ n\right] !}\text{ .}  \label{24}
\end{equation}%
Using the definition of $q$-derivative $\left( 18\right) $, one can derive
the following expression\ \ \ \ \ \ \ \ \ \ 
\begin{equation}
\dfrac{dE\left( q,\lambda ;x\right) }{d_{q}x}=E\left( q,\lambda ;q^{2\lambda
}x\right)  \label{25}
\end{equation}%
so with the parts formula$\left( 22\right) $ and $\left( 23\right) $, the
following result can be carefully calculated.\ \ \ \ \ \ \ 
\begin{equation}
\ \int_{0}^{\infty }E\left( q,\lambda ;-x\right) x^{n}d_{q}x=q^{h\left(
n,\lambda \right) }\left[ n\right] !\text{ ,}  \label{26}
\end{equation}%
where

\begin{equation}
h\left( x,\alpha \right) =\alpha \left( x+1\right) \left( x+2\right)
-x\left( x+1\right) /2\text{ .}  \label{27}
\end{equation}%
Next step we define $q\left( \lambda ,\rho \right) $-modified Bessel
function which has the same form as the common modified Bessel function when 
$q,\lambda ,\rho $ approach $1$ simultaneously\ \ \ \ \ \ \ \ \ \ \ \ \ \ \
\ \ 
\begin{equation}
K_{n}\left( \lambda ,\rho ;x\right) =\frac{1}{\left[ 2\right] _{q}}\left( 
\dfrac{x}{\left[ 2\right] _{q^{1/2}}}\right) ^{n}\int_{0}^{\infty
}t^{-n-1}E\left( q,\lambda ;-t\right) E\left( q,\rho ;-\frac{x^{2}}{\left[ 2%
\right] _{q^{1/2}}^{2}t}\right) d_{q}t\   \label{28}
\end{equation}%
\ where $n$ is integer. With the above definition and $\left( 26\right) $,
we have

\begin{eqnarray}
&&\int_{0}^{\infty }d_{q^{1/2}}r\cdot r^{\mu }K_{\nu }\left( \lambda ,\rho ; 
\left[ a\right] _{q^{1/2}}r\right)  \notag \\
&=&q^{h\left( \frac{\mu +\nu -1}{2},\rho \right) +h\left( \frac{\mu -\nu -1}{%
2},\lambda \right) }\left[ 2\right] _{q^{1/2}}^{\mu -1}\left[ a\right]
_{q^{1/2}}^{-\mu -1}\Gamma _{q}\left( \tfrac{\mu +\nu +1}{2}\right) \Gamma
_{q}\left( \tfrac{\mu -\nu +1}{2}\right)  \label{29}
\end{eqnarray}%
where $\Gamma _{q}\left( x\right) $ is $q$-deformed Gamma Function. When $q$
approaches $1$, it reduces into the common Gamma Function. If $x$ is
integer, it equals $\left[ x\right] !$. By setting%
\begin{equation}
\Phi _{e}\left( z\right) =q^{e+3}\left[ 2\right] _{q^{1/2}}\left\vert
z\right\vert ^{e}N_{e}^{-2}K_{e}\left( \lambda ,\rho ;\left[ 2\right]
_{q^{1/2}}\left\vert z\right\vert \right)  \label{30}
\end{equation}%
\ where%
\begin{equation*}
\ \lambda =1+2e^{-1},\ \ \ \ \rho =1/2-2e^{-1}
\end{equation*}%
these charged coherent states for the two-mode coupled $q$-bosons satisfy
the completeness relation%
\begin{equation}
\overset{\infty }{\underset{e=-\infty }{\sum }}\iint \dfrac{d_{q}^{2}z}{\pi }%
\Phi _{e}\left( z\right) \left\vert z,e\right\rangle \left\langle
z,e\right\vert =I\ \ \ \ \   \label{31}
\end{equation}%
where

\begin{equation*}
d_{q}^{2}z\equiv \tfrac{1}{2}d_{q}\left( \left\vert z\right\vert ^{2}\right)
d\theta =\tfrac{\left[ 2\right] _{q^{1/2}}}{2}\left\vert z\right\vert
d_{q^{1/2}}\left( \left\vert z\right\vert \right) d\theta \text{ .}
\end{equation*}%
\ \ \ \ \ \ \ \ \ \ \ \ \ \ \ \ \ \ \ \ \ \ \ \ \ \ \ \ \ \ \ \ \ \ \ \ \ \
\ \ \ \ \ \ \ \ \ \ \ \ \ \ \ \ \ \ \ \ \ \ 

The coherent states for this system may also be obtained by projecting out a
state of a definite charge from the following state\ \ \ \ \ \ \ \ \ 
\begin{eqnarray}
\left\vert \alpha \beta \right\rangle &=&\overset{\infty }{\underset{n.m=0}{%
\sum }}\dfrac{\alpha ^{n}\beta ^{m}}{q^{m\left( m-1\right) /4}\left[ n\right]
!\left[ m\right] !}\left( b^{\dag }\right) ^{m}\left( a^{\dag }\right)
^{n}\left\vert 0,0\right\rangle  \notag \\
&=&\overset{\infty }{\underset{n.m=0}{\sum }}\dfrac{\alpha ^{n}\beta ^{m}}{%
\sqrt{q^{m\left( m-1\right) /2}\left[ n\right] !\left[ m\right] !}}%
\left\vert n,m\right\rangle \ \   \label{32}
\end{eqnarray}%
which does not have a definite charge. It is easily found that this state is
the associated coherent state of $ab$ and $ba$ satisfying%
\begin{equation}
ab\left\vert \alpha \beta \right\rangle =\alpha \beta \left\vert \alpha
\beta \right\rangle ,\text{ \ \ \ }ba\left\vert \alpha \beta \right\rangle =%
\sqrt{q}\alpha \beta \left\vert \alpha \beta \right\rangle  \label{33}
\end{equation}%
If we set\ \ \ \ \ \ \ \ \ 
\begin{equation}
\alpha =\xi Exp\left( -i\left( \theta +\varphi \right) \right) ,\ \ \ \
\beta =\eta Exp\left( -i\left( \theta -\varphi \right) \right) ,  \label{34}
\end{equation}%
we can find that the state\ \ \ \ \ \ \ \ 
\begin{equation}
\ \left\vert \xi \eta \theta ;e\right\rangle =\frac{1}{2\pi }\int_{0}^{2\pi
}d\varphi Exp\left( ie\varphi \right) \left\vert \alpha \beta \right\rangle
\label{35}
\end{equation}%
yields\ \ \ \ \ \ \ 
\begin{equation}
Q\left\vert \xi \eta \theta ;e\right\rangle =e\left\vert \xi \eta \theta
;e\right\rangle ,\ \ \ \ \ \ \ \ \ \ ab\left\vert \xi \eta \theta
;e\right\rangle =\alpha \beta \left\vert \xi \eta \theta ;e\right\rangle ,\ 
\label{36}
\end{equation}%
so this state is identical to the charged coherent state $\left\vert
z,e\right\rangle $ for this coupled $q$-boson system if we only set $\xi
\eta Exp(-i\theta )=z$. Here $Exp(x)$ is the ordinary exponential function
of $x$.

In summary, we explicitly construct the $q$-deformed charged coherent states
for the the two-mode coupled $q$-bosons with $su_{q}(2)$ covariance and give
a resolution of unity for these states. We also find a simple way to obtain
these coherent states using state projection. Similarly one can construct
deformation of the charged coherent states for multimode coupled $q$-bosons$%
(1)$. Work on this direction is in progress.\ \ \ \-

\end{document}